\documentstyle[aps,epsfig]{revtex}
\def\be{\begin{eqnarray}}
\def\ee{\end{eqnarray}}

\begin{document}
\twocolumn[\hsize\textwidth\columnwidth\hsize\csname %
@twocolumnfalse\endcsname
\tighten
\title{Thermodynamics of $n$-$p$ condensate in asymmetric nuclear matter}
\author{ A. Sedrakian$^{1)}$ and U. Lombardo$^{2)}$} 
\address{ $^{1)}$Kernfysisch Versneller Instituut,
         NL-9747 AA Groningen,
         The Netherlands\\
          $^{2)}$Dipartimento di Fisica and INFN, 57 Corso Italia, 
         I-95129 Catania, Italy} 
\maketitle
\begin{abstract}
We study the neutron-proton pairing in nuclear matter
as a function of isospin asymmetry at finite 
temperatures and the saturation density using  
realistic nuclear forces and Brueckner-renormalized single particle 
spectra.  Our computation of the thermodynamic
quantities shows that while the difference of the entropies of the 
superconducting and normal phases anomalously changes its 
sign as a function of  temperature for arbitrary asymmetry,
the grand canonical potential does not; the superconducting state
is found to be stable in the whole 
temperature-asymmetry plane. The pairing gap completely disappears  
for density-asymmetries exceeding $\alpha_c= (n_n-n_p)/n \simeq 0.11$.
\end{abstract} 
]

\date{today}

Neutron-proton pair correlations are potentially 
important in a number of contexts, including the 
study of the nuclear structure of  medium mass
$N\simeq Z$ nuclei produced at the radioactive nuclear 
beam facilities\cite{AG99} and the theory of the 
deuteron formation in the medium energy 
heavy-ion collisions\cite{Baldo2}. 
In the astrophysical context
$n$-$p$ pairing correlations are relevant for 
the astrophysical $r$-process\cite{Kr93,Che95} 
and could could play a major role in neutron 
star models which permit pion or 
kaon condensation\cite{Bethe94}.

In recent years much theoretical effort has been devoted to the 
understanding of the $n$-$p$ pair correlations. They have been studied
both in the infinite 
nuclear matter within the Thouless \cite{TH,Eme60} 
or the Bardeen-Cooper-Schrieffer \cite{FW} theories of superconductivity
\cite{ALM1,VONDER,BALDO1,ALM2,ALM3,SAL97,BALDO3,MORTEN}, 
and  in finite nuclei within mean-field effective interaction 
theories of 
pairing\cite{AG99,Engel96,Engel97,Doba96,Satula97,Civ97,Civ97_bis,Roepke00}. 
In particular,  microscopic calculations, based on the BCS 
theory for the bulk nuclear matter
show that the isospin-asymmetric matter supports Cooper-type 
pair correlations in the $^3S_1$-$^3D_1$ partial-wave channel 
due to the tensor component of the nuclear force. The energy gap
is of the order of 10 MeV at the saturation 
density\cite{ALM1,VONDER,BALDO1,MORTEN} when
the effects of the medium polarization on the pairing 
force\cite{CLARK1,CLARK2,NISKANEN,WAMBACH,SCHULZE} are 
neglected.

As the existence of the pair correlations crucially depends upon
the overlap between the neutron and proton Fermi surfaces,  
one expects a suppression of the pairing correlations
if the system is driven out of the isospin-symmetric state.
The studies based on the 
Thouless criterion\cite{TH,Eme60} for the thermodynamic $T$-matrix
(the divergence of the of ladder resummation scheme 
at the critical temperature) deduced  the 
suppression of the critical temperature with the isospin 
asymmetry\cite{ALM2,ALM3}; however, they do not permit to 
draw conclusions about properties of the superconducting state
with a finite gap. The studies based on the  
BCS theory without self-energy-corrections to 
the single particle spectrum\cite{SAL97,BALDO3}, 
while give a correct qualitative picture,
overestimate the magnitude of the pairing gap
and the critical asymmetries at which the pairing disappears.

In this paper we report on a fully microscopic calculation of the 
$n$-$p$ pairing within the strong coupling theory of the 
superconductivity with realistic nuclear interactions. 
Our main focus is the effect of 
ladder-renormalized single particle
energies\cite{ZUO} on the magnitude of the pairing 
gap and the threshold asymmetries
at which the pairing vanishes\cite{REMARK}. 
Our second aim is to provide a first 
computation of the key  thermodynamic quantities 
of asymmetric $n$-$p$ condensate from the strong 
coupling BCS theory.

To set the stage, let us start with the solution of
the Dyson equations for the normal and anomalous propagators
in the  Matsubara formalism. 
In doing so we shall decouple isospin singlet 
$SD$ pairing channel from the isospin triple channels,
as the $SD$ coupled channels contain the dominant part of 
the attractive pairing force. In this 
case the pairing matrix is diagonal in the spin-space, i.e. 
one deals with the unitary triplet state (see ref. \cite{BALDO1}). 
The proton/neutron propagators follow from the  
solution of the Gor'kov equations, and can be cast 
in the form  $(\hbar = 1)$
\be\label{G1}
G^{(p/n)}_{\sigma, \sigma^\prime}({\vec k},\omega_m) \, 
= {\,- \delta_{\sigma,\sigma^
\prime} \, {i\omega_m +\xi_k \mp \delta \varepsilon_{\vec k}\over
{(i\omega_m + E_{\vec k}^+)(i\omega_m - E_{\vec k}^-)}}} \, ;
\ee
the neutron-proton anomalous propagator has the form  
\be\label{G2}
F^{\dagger}_{\sigma, \sigma'}({\vec k},\omega_m) =
- \delta_{\sigma,\sigma^\prime} \frac{\Delta^{\dagger}
(\vec k) }
{(i\omega_m + E_{\vec k}^+)(i\omega_m - E_{\vec k}^-)} \, ,
\ee
where $\omega_m$ are the Matsubara frequencies, 
the upper sign in $G^{(p/n)}$ corresponds to 
protons and the lower to neutrons.
The isospin asymmetry lifts the  degeneracy of the quasiparticle 
spectra thus leading to   two separate branches for  
protons and neutrons
\be\label{SPS}
E_{\vec k}^{\pm} &=& \sqrt{\xi^2_{\vec k}+\Delta_{\vec k}^2}  
\pm\delta \varepsilon_{\vec k}\, ,
\ee
where
\be
\xi_{\vec k} & \equiv &\frac{1}{2}\left( \varepsilon_{\vec k}^{(n)}+ \varepsilon_{\vec k}^{(p)} \right), \qquad 
\delta \varepsilon_{\vec k}\equiv \frac{1}{2} \left(\varepsilon^{(p)}_{\vec k} 
- \varepsilon^{(n)}_{\vec k} \right)\nonumber ,
\ee
and  $\varepsilon_{\vec k}^{(n,p)}$ are the single  
particle energies of neutrons 
and protons. The strong coupling BCS theory is coupled to the 
Brueckner renormalizaton scheme via the 
single particle energies defined as 
$\varepsilon_{\vec k}^{(n,p)}= k^2/2m + {U^{(n,p)}(k)}-\mu^{(n,p)}$;
here  $U^{(n,p)}(k)$ are the single particle potentials  
which are derived from the Brueckner 
theory of asymmetric nuclear matter \cite{ZUO}
and $\mu^{(n,p)}$ are the chemical potentials for neutrons and protons,
which are derived from the BCS theory self-consistently. 
Both schemes are normalized to the same densities. The 
small effect of the feedback of the pairing correlation in 
the Brueckner calculations of the mean-field\cite{LOMBARDO} 
is, however, neglected.

Using the angle-averaging procedure, which is an adequate
approximation for the present purpose (see ref. \cite{BALDO1})
the BCS gap equation for asymmetric 
nuclear matter can be derived. We find
\be\label{GAP}
 \Delta_l(k)\, &=&\, -\sum_{k^\prime}\sum_{l^\prime}
  V_{l l^\prime}(k,k^\prime)\nonumber\\
&\times& {\Delta_{l^\prime}(k^\prime)\over 
  2\sqrt{{\xi}_{k^\prime}^2 + D(k^\prime)^2} }
[1-f({E_{\vec k}}^{+})-{f(E_{\vec
 k}}^{-})],
\ee
where $D(k)^2\equiv\Delta_0(k)^2+\Delta_2(k)^2$ is 
the angle-averaged neutron--proton 
gap function, $f(E)=[1+\exp(\beta E)]^{-1}$ is the 
Fermi distribution function,
$\beta^{-1}=k_BT$, where $T$ is the 
temperature and $k_B$ is the Boltzmann constant. 
The driving term,  $V_{l l^\prime}$, 
is the bare interaction in the $SD$ channel. 
The density matrices of neutrons and protons 
follow from eq. (\ref{G1}) after summation  over frequencies,
\be\label{DENS}
n^{(p/n)}_{\sigma}( k)
&=& \Biggl\{ \frac{1}{2}\Bigl(1+\frac{\xi_{\vec k}}{\sqrt{\xi^2_{\vec k}
+\Delta_{\vec k}^2}}\Bigr) f(E_{\vec k}^{\pm}) \nonumber \\
&+& \frac{1}{2}\Bigl(1-\frac{\xi_{\vec k}}{\sqrt{\xi^2_{\vec k}+\Delta_{\vec k}^2}}\Bigr)
\left[1- f(E_{\vec k}^{\mp})\right]\Biggr\}.
\ee 
Summation  over frequencies in eq. (\ref{G2})
leads to the density matrix of the particles in the condensate 
\be
\nu (\vec k)=  \frac{\Delta(\vec k)}{2\sqrt{\xi^2_{\vec k}
+\Delta_{\vec k}^2}}   
[1-f({E_{\vec k}}^{+})-{f(E_{\vec k}}^{-})].
\ee
The partial densities of nucleons, in terms of the density matrices, 
are $\rho^{(p/n)} = \sum_{{\vec k},\sigma}n^{(p/n)}_{\sigma}(k)$; 
similarly,
$\rho^{\rm cond} = \sum_{{\vec k},\sigma}\nu_{\sigma}(k)$ is the
the density of the particles within the condensate.
It is essential that
the coupled system of equations (\ref{GAP}) and (\ref{DENS}) 
are solved self-consistently.

Now we are in the position to write down the key thermodynamic 
quantities. As the occupation of the quasiparticle states is 
given by the Fermi-Dirac distribution function, the entropy 
of the system is given by the combinatorical expression
\be\label{ENTROPY}
S &=& - 2 k_B \sum{\vec k} 
\Bigl\{f({E_{\vec k}}^{+})\,{\rm ln}\, f({E_{\vec k}}^{+})+
        \bar f({E_{\vec k}}^{+})\,{\rm ln}\, 
\bar f({E_{\vec k}}^{+})\nonumber\\
&+&f({E_{\vec k}}^{-})\,{\rm ln}\, f({E_{\vec k}}^{-}) 
+\bar f({E_{\vec k}}^{-})\,{\rm ln}\, \bar f({E_{\vec k}}^{-}) 
\Bigr\},
\ee
where $\bar f({E_{\vec k}}^{\pm}) = [1-f({E_{\vec k}}^{\pm})]$.
Equation (\ref{ENTROPY}) is an obvious extension of the 
mean-field expression for the entropy in the symmetrical nuclear 
matter\cite{BALDO1} to the asymmetric case.
The internal energy, defined as the grand canonical
statistical average of the Hamiltonian, 
${\cal U} = \langle \hat H-\mu^{(n)}
\hat\rho_n - \mu^{(p)}\hat\rho_p\rangle$, reads
\be\label{INT_E} 
{\cal U}(T,\mu) &=& \sum_{\sigma\vec k} \left[\varepsilon_{\vec k}^{(n)}
n^{(n)}_{\sigma}( k)
+\varepsilon_{\vec k}^{(p)}n^{(p)}_{\sigma}( k)\right]\nonumber \\
 &+& \sum_{\vec k_1\vec k_2}
\, V(\vec k_1,-\vec k_1; \vec k_2,-\vec k_2)\, 
\nu(\vec k_1)\nu(\vec k_2), 
\ee
where we have carried out the spin summation in the second term.
The non-pairing interaction energy among the quasiparticles and the 
contribution from the chemical potentials are included in 
the single particle energies (the first term) of eq. (\ref{INT_E}). 
The second term includes the BCS mean 
field interaction among the particles in the condensate.
Note that the interaction in the second term 
of eq. (\ref{INT_E}) can be eliminated in terms of the gap equation
(\ref{GAP}).  Finally, the thermodynamic potential is given as
\be\label{GR} 
\Omega(T,\mu) = {\cal U} (T,\mu) - TS.
\ee
The  expressions for the entropy and the 
internal energy in the normal state
follow from eqs. (\ref{ENTROPY}) and 
(\ref{INT_E}) in the limit $\Delta= 0$. Our 
main interest below is  the change 
in the thermodynamic potential of the superconducting 
state with respect to the normal state 
$\delta\Omega = \Omega_s -\Omega_n$.

Numerical calculations of the pairing gap 
were carried out using the Paris potential. The single particle 
potentials were used as an input from Brueckner computation 
based on the Argonne V14 potential; for details of these
calculations we refer to ref. \cite{ZUO}.

Fig. 1 shows the values of the pairing gap in the $^3S_1$-$^3D_1$ 
partial wave channel as a function of the temperature  
at the saturation density  $\rho = 0.17$ fm$^{-3}$.
The asymmetry parameter is defined as 
$\alpha \equiv (\rho_n-\rho_p)/\rho$. For the value 
$\alpha = 0$ one finds the usual BCS solution: the gap is 
a monotonically decreasing function of the temperature and 
vanishes at the critical temperature $T_c = \Delta(T=0)/ 1.76$. 
In the asymmetric state the gap develops a maximum at a certain
intermediate temperature in the range $0< T < T_c$; the value of the critical 
temperature is reduced and the BCS relation between the critical 
temperature and the value of the gap at $T=0$ does not hold any longer.
For large asymmetries $\alpha \sim 0.1$ the superconducting state
exists only in a finite temperature range; it completely vanishes 
for the critical asymmetry $\alpha_c = 0.11$.   
This behavior is the result of the interplay between the increasing 
shift between the radii of the Fermi spheres of neutrons and protons 
with increasing asymmetry
and the smearing of the Fermi surfaces due to {\it both} interaction 
driven correlations and the temperature. 
While the first factor tends 
to suppress the pairing, the role of the
second factor is twofold. 
On the one hand the temperature and correlations
smear the Fermi surfaces thus increasing 
the overlap between the two Fermi spheres which  
promotes the pairing. 
If, however, the temperatures are high enough the 
pair correlated states are quenched by the 
thermal excitation. 
The pairing gap, hence, has a maximum at 
some intermediate temperature. The two critical 
temperatures are controlled by two different mechanisms. 
The superconductivity vanishes with decreasing 
temperature at a lower critical 
temperature when the smearing becomes 
insufficient to support the pairing. The   
superconductivity vanishes at the upper critical temperature 
because of the thermal excitation of the system, as in the 
standard BCS theory.

Qualitatively, the role of the mean-field can be understood in 
terms of the effective mass.
As in the symmetric matter the nucleon effective mass, 
$m^*$, is reduced by the 
mean-field, the density of states at the Fermi surface is reduced, 
and therefore the pairing gap is reduced as well. 
In asymmetric matter there are no additional 
non-trivial effects. 
Even though the neutron and proton effective masses change 
with the denisty in
opposite directions, the gap itself depends on the 
reduced effective mass $ (m^*_p+m^*_n)/2m^*_p m^*_n$, which 
scales with the asymmetry as $1/m^* +{\cal O}(\alpha^2)$.

Fig. 2 displays the entropies of the normal and superconducting 
states as a function of the temperature  at the saturation 
density. The $\alpha = 0$ behavior of the normal and superconducting 
entropies is the one expected from the 
theory of the normal Fermi-liquids and the BCS theory, respectively:
the entropy of the normal state is a linear function of the 
temperature; the entropy of the superconducting state is linear 
close to the critical temperature and  decreases
exponentially in the low temperature limit. For finite 
asymmetries the entropy of the normal state is
unchanged. The curve of the entropy of superconducting state, however, 
shows an anomalous behaviour by crossing the entropy curve of 
the normal state. For temperatures below the crossing point
{\it the entropy of the superconducting state is larger 
than that of the normal state}. The crossover to the anomalous 
behaviour occurs at the temperature at which the gap reaches its
maximum. This anomalous dependence of the entropy of the 
superconducting state on the temperature implies
that this state could be unstable for 
temperatures below the temperature at which the gap attains 
its maximum. The internal energy, the first term  
in eq. (\ref{GR}),  has a definite
sign as it vanishes in the normal state and 
is a quadratic form of the pairing gap.  Thus the sufficient 
condition for the onset of the instability is  
that the  contribution of the second term in 
eq. (\ref{GR}) dominates that of the first
in the range of the temperatures below the crossover point.

Fig. 3 shows the difference in the thermodynamic potentials
of the normal and superconducting state  $\delta\Omega$. This 
quantity is negative for all temperatures and asymmetries.
The anomalous change of sign of $S_n-S_s$, therefore,
does not change the net balance between the energies of the 
normal and superconducting states. The temperature dependence 
of $\delta\Omega$ is 
the reminiscent of the tempertaure dependence of the gap function
by virtue of the 
(approximately) quadratic dependence of the internal energy 
on the pairing gap. 
We thus conclude that superconducting state is  stable, at 
least in the present model, in the whole 
temperature-asymmetry plane, whenever a non-trivial 
solution to the gap equation exists. 
We have verified that the deduced behaviour of 
the $\delta\Omega$ is the consequence
of the fact that the contribution of the
internal energy to the thermodynamic potential  
dominates the contribution from the entropy.

The problem of the pairing suppression and the thermodynamics 
of the superconducting phase when the particles
lie on different Fermi-surfaces 
is encountered in many systems. 
The simplest example is a superconducting
metal in a magnetic field, where the spin degeneracy is 
relaxed because of the Pauli paramagnetism\cite{LARKIN,FULDE,LEGGET}. 
The  spin zero pairing between the spin-up and spin-down populations
is precisely the one described above, albite one commonly 
uses the BCS weak coupling ansatz which does not work for
nuclear matter.  Another example is  
the B-state of liquid $^3$He (Balian-Werthamer state)
in the magnetic field\cite{LEGGET2}. 
In both cases  measurements of the specific heat and spin 
susceptibility  could provide an experimental information on 
the anomalous behaviour of the entropy. 
Apart from condensed matter, our
results may be relevant for the current
discussion of spontaneous braking of color symmetry 
in finite temperature/density QCD
and formation of  $\langle qq \rangle$ color superconducting
 condensate\cite{ALFORD,RAPP,BLASCHKEWICZ,PISARSKI}. 
In all cases, it would be interesting to check whether the 
balance between the internal energy and the 
entropy is generically such that the superconducing state 
remains stable for arbitrary asymmetries. 
 
To conclude, we studied the pairing in the isospin asymmetric 
nuclear matter using realistic nuclear interactions combined with 
ladder-renormalized single particle energies. The pairing 
at the saturation density vanishes for the critical density 
asymmetry $\alpha_c = 0.11$ at the finite temperature $T\simeq 1.5$
MeV. Our evaluation of the thermodynamic quantities of the 
isospin asymmetric nuclear matter shows that,
while the entropy of the superconducting state becomes larger 
than that of the normal state below a certain temperature, 
the thermodynamical potential at the saturation density
corresponds to a stable superconducting state 
in the whole asymmetry-temperature plane. 

\section*{Acknowledgements}
We acknowledge discussions and correspondence 
with Professors  P. Nozi\`eres, G. R\"opke and 
P. Schuck. AS thanks the LNS, INFN  for the warm hospitality
during his stay in Catania. The work at KVI, Groningen,
has been supported through the Stichting voor Fundamenteel 
Onderzoek der Materie with financial support from the 
Nederlandse Organisatie voor Wetenschappelijk Onderzoek.

\newpage

\begin{figure}
\begin{center}
\mbox{\epsfig{figure=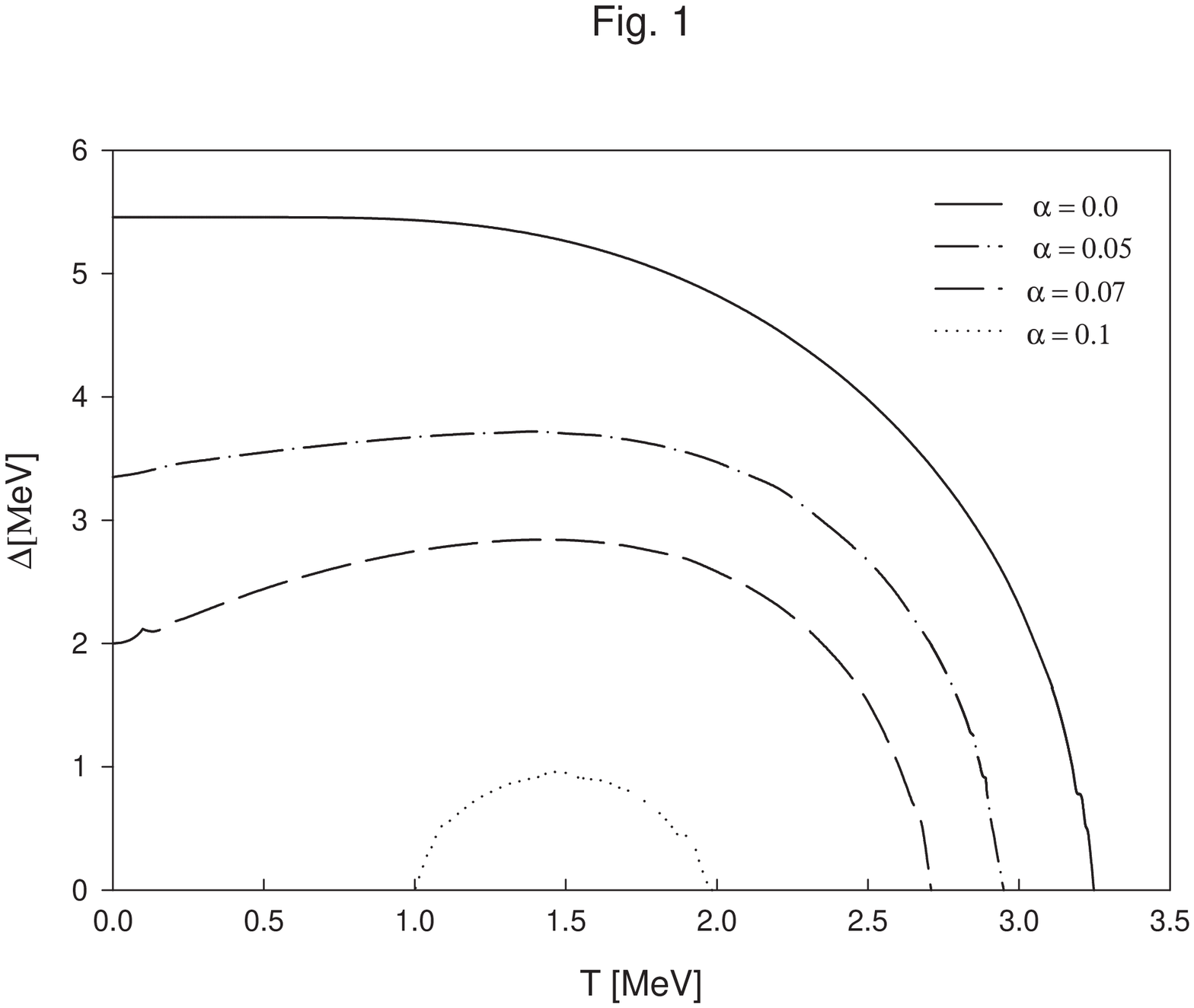,height=10.in,width=6.in,angle=0}}
\end{center}
\caption[]
{\footnotesize{The pairing gap at the saturation density 
as a function of the temperature for asymmetries 
$\alpha = 0$, 0.05, 0.07 and  0.1.}}
\label{fig1}
\end{figure}

\newpage
\begin{figure}
\begin{center}
\mbox{\epsfig{figure=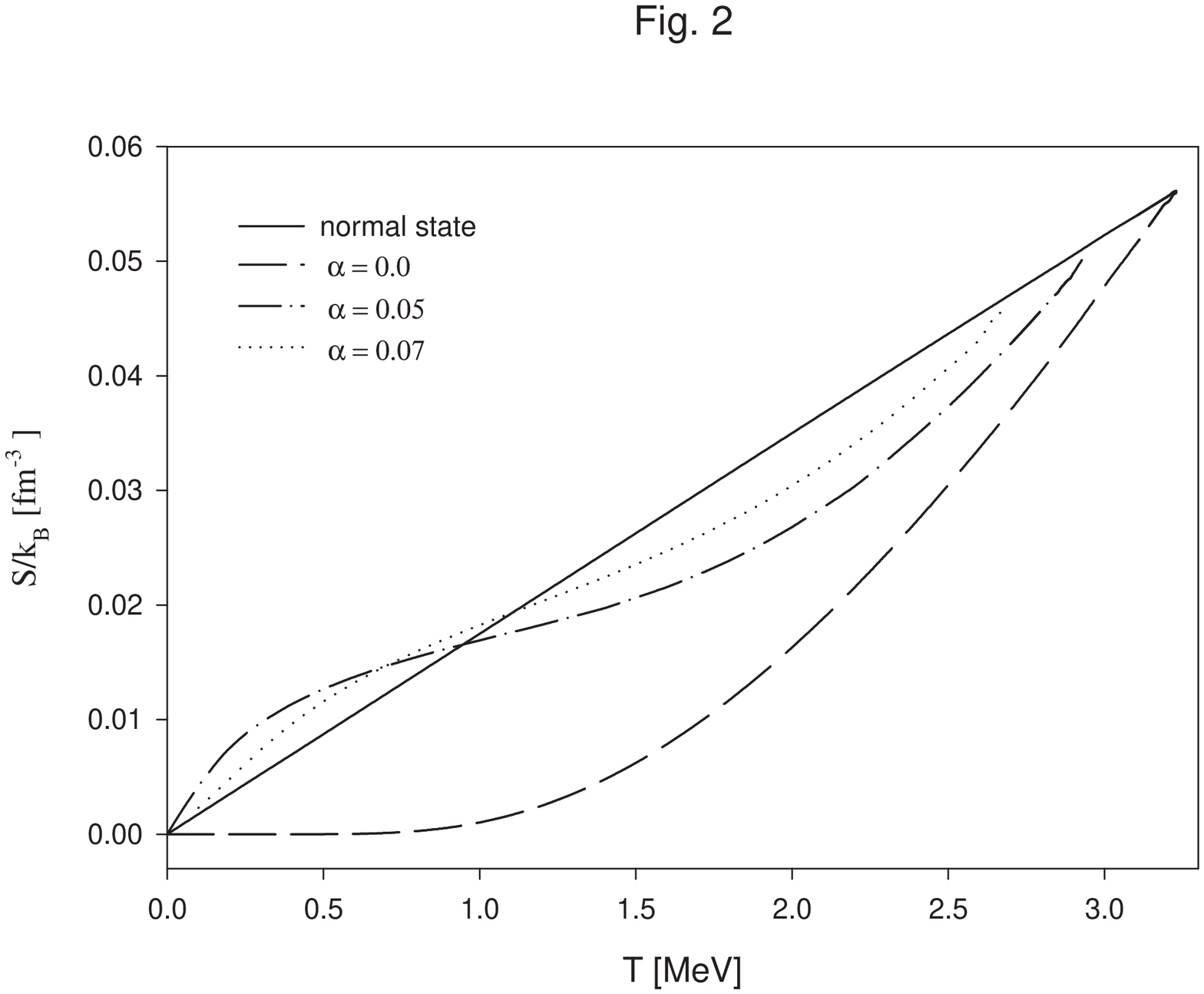,height=10.in,width=6.in,angle=0}}
\end{center}
\caption[]
{\footnotesize{The entropy of normal (solid line) and superconducting
state for $\alpha = 0$ and  0.05, 0.07; the $\alpha = 0.1 $ curve is 
indistinguishable from the normal state entropy curve on figure's scale.}}
\label{fig2}
\end{figure}

\newpage
\begin{figure}
\begin{center}
\mbox{\epsfig{figure=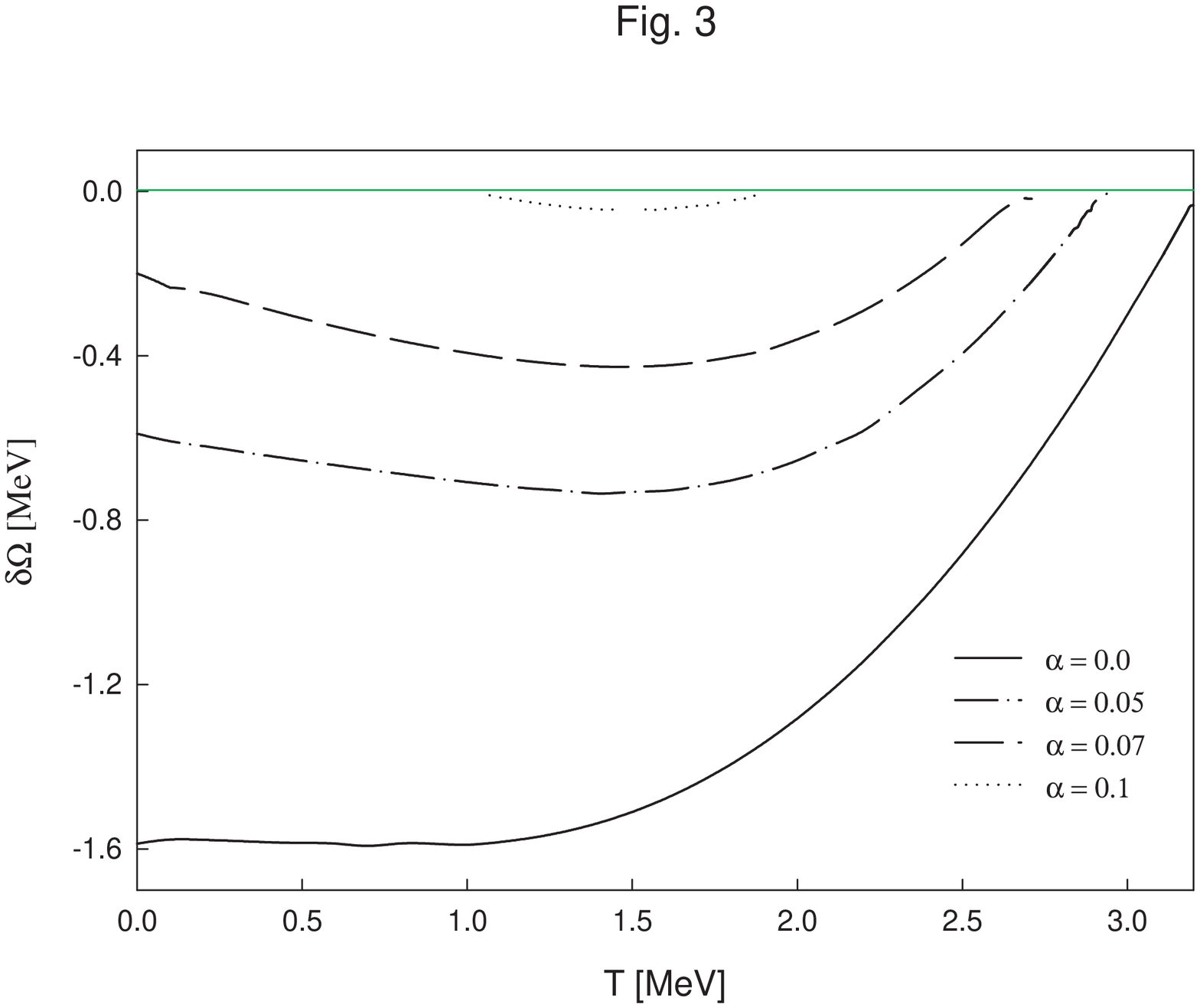,height=10.in,width=6.in,angle=0}}
\end{center}
\caption[]
{\footnotesize{The difference in the thermodynamic potentials 
of the superconducting and normal states  at the saturation density 
as a function of the temperature for asymmetries 
$\alpha = 0$, 0.05, 0.07 and  0.1.}}
\label{fig1}
\end{figure}


\begin{thebibliography}{99}


\bibitem{AG99}  Alan L. Goodman, Phys. Rev. {\bf C60}, 014331 (1999),
                and references therein.


\bibitem{Baldo2} M. Baldo, U. Lombardo, and P. Schuck, 
                 Phys. Rev.  {\bf C52},  975  (1995).


\bibitem{Kr93}B.  K.-L. Kratz, J.-P. Bitouzet, F.-K. Thielmann,  and
              B. Pfeiffer, Astrophys. J. {\bf 403}, 216 (1993).

\bibitem{Che95} B. Chen, J. Dobaczewski, K.-L. Kratz, K. Langanke, B. Pfeiffer, F.-K.
                Thielmann, and P. Vogel, Phys. Lett. {\bf B355}, 37 (1995).


\bibitem{Bethe94} G. E. Brown and H. A. Bethe, ApJ {\bf 423 }, 659 (1994). 


\bibitem{TH} D. J. Thouless, Ann. Phys. {\bf 10},  553 (1960).

\bibitem{Eme60}V. J. Emery and A. M. Sessler, Phys. Rev. {\bf 119}, 248 (1960).

\bibitem{FW} A. L. Fetter and J. D. Walecka, {\it Quantum 
            Theory of Many-Particle Systems} (McGraw-Hill, New York, 1971).

\bibitem{ALM1} T. Alm, G. R\"opke, 
              and M. Schmidt, Z. Phys. {\bf A337},  355 (1990).
              

\bibitem{VONDER}  B. E. Vonderfecht, C. C. Gearhart, W. H. Dickhoff, 
                  A. Polls, and  A. Ramos, Phys. Lett. {\bf B253},  1 (1991).

\bibitem{BALDO1}  M. Baldo, I. Bombaci and U. Lombardo, Phys. Lett. 
                  {\bf B283}, 8 (1992).

\bibitem{ALM2}  T. Alm, B. L. Friman, G. R\"opke, and H. Schulz,  Nucl. Phys. 
               {\bf A 551}, 45 (1993).

\bibitem{ALM3}  T. Alm, G. R\"opke, A. Sedrakian, and F. Weber, 
                Nucl. Phys. {\bf A406},  491 (1996).

\bibitem{SAL97} A. Sedrakian, T. Alm, 
                and U. Lombardo, Phys. Rev. {\bf C55}, R582 (1997).

\bibitem{BALDO3} M. Baldo, U. Lombardo, P. Schuck, and A. Sedrakian, 
                Condensed Matter Theories Vol. 12 (1997) editor J. W. Clark,
                Nova Science Publishers, pg. 265-277.


\bibitem{MORTEN} Oe. Elgaroy, L. Engvik, E. Osnes, and M. Hjorth-Jensen, 
                 preprint nucl-th/9709039.


\bibitem{Engel96} J. Engel, K. Langanke, and P. Vogel, Phys. Lett. B 
                  {\bf 389}, 211 (1996). 



\bibitem{Engel97} J. Engel, S. Pittel, M. Stoitsov, P. Vogel, 
                  and J. Deukelsky, Phys. Rev.  {\bf C55}, 1781 (1997).


\bibitem{Doba96} J. Dobaczewski, W. Nazarewicz, T.R. Werner, J. F. Berger,
                 C. R. Chinn, and J. Decharge, Phys. Rev. {\bf C53}, 2809 
                 (1996).
\bibitem{Satula97} W. Satula and R. Wyss, Phys. Lett. B {\bf 393}, 
                   1 (1997).

\bibitem{Civ97} O. Civitarese, M. Reboiro, and P. Vogel, Phys. Rev.  
                  {\bf C56}, 1840 (1997). 

\bibitem{Civ97_bis} O. Civitarese and M. Reboiro, Phys. Rev.  
                  {\bf C56}, 1179 (1997).                

\bibitem{Roepke00} G. R\"opke, A. Schnell, P. Schuck, and U. Lombardo, 
                 preprint nucl-th/9906092.


\bibitem{CLARK1} J. W. Clark, C. G. K\"allman, C. H. Yang, and D. A. Chakkalakal, 
                 Phys. Lett. B {\bf 61}, 331 (1976).

\bibitem{CLARK2}  J. M. C. Chen, J. W. Clark, R, D. Dav\'e, and V. V. Khodel, 
                  Nucl. Phys. A {\bf 555}, 59 (1993).  
\bibitem{NISKANEN} J. A. Niskanen and J. A. Sauls, unpublished (1981).

\bibitem{WAMBACH} J. Wambach, T. L. Ainsworth, and D. Pines, Nucl. Phys. A
                  {\bf 555}, 128 (1993).
\bibitem{SCHULZE} H.-J. Schulze, J. Cugnon, A. Lejeune, M. Baldo, 
                 and U. Lombardo, Phys. Lett. {\bf B375}, 1 (1996).



\bibitem{ZUO} W. Zuo, I. Bombaci, and U. Lombardo, Phys. Rev. C, in press.


\bibitem{REMARK} We do not discuss the separate issue of the influence 
                 of the medium polarization on the pairing force, an
                 issue of current debate.
                 One should bare, however, in mind that the calculations of the 
                 $^1S_0$ pairing in neutron 
                 matter \cite{CLARK1,CLARK2,NISKANEN,WAMBACH,SCHULZE} indicate considerable 
                 modifications in the values of the pairing gaps
                 due to the medium polarization.
\bibitem{LOMBARDO} U. Lombardo, H.-J. Schulze, and W. Zuo, Phys. Rev C {\bf
59}, 2927, (1999).
\bibitem{LARKIN} A. I. Larkin and Yu. N. Ovchinnikov, Sov. Phys. JETP 
                 {\bf 20},  762 (1965).
\bibitem{FULDE} P. Fulde and R. A. Ferrel, Phys. Rev. {\bf 135}, 550, (1964).

\bibitem{LEGGET} A. Legget, Phys. Rev. {\bf 140}, A1869 (1965).

\bibitem{LEGGET2} A. Legget, Rev. Mod. Phys. {\bf 41}, 331 (1975).

\bibitem{ALFORD} M. Alford, K. Rajagopal, and F. Wilczek,
                 Phys. Lett. {\bf B422},  247 (1998).
\bibitem{RAPP}R. Rapp, T. Schaefer, E.V. Shuryak, and  M. Velkovsky,
                 Phys. Rev. Lett. {\bf 81}, 53 (1998).
\bibitem{BLASCHKEWICZ} D. Blaschke and C. D. Roberts, Nucl. Phys. A 
                       {\bf 642}, 197 (1998).
\bibitem{PISARSKI} R. Pisarski and D. Rischke, preprint nucl-th/9903023.

\end{thebibliography}
\end{document}